\documentclass[conference]{IEEEtran}
\IEEEoverridecommandlockouts
\usepackage{amsmath,amssymb,amsfonts}
\usepackage{algorithmic}
\usepackage{graphicx}
\usepackage{textcomp}
\usepackage{xcolor}
\usepackage{tikz}
\usepackage[sort]{cite}
\usepackage[hidelinks]{hyperref}
\usepackage{booktabs}
\usepackage{flushend}
\usepackage{amssymb}
\usetikzlibrary{shapes,arrows}
\usepackage{psfrag}
\usepackage{makecell}
\usepackage{subcaption}

\makeatletter
\def\blfootnote{\xdef\@thefnmark{}\@footnotetext}
\renewcommand\footnoterule{\kern-3\p@ \hrule \@width 2in \kern 2.6\p@} % Adjust rule length here
\makeatother
    
\begin{document}

\title{Align-ULCNet: Towards Low-Complexity and Robust Acoustic Echo and Noise Reduction\
}

\author{\IEEEauthorblockN{Shrishti Saha Shetu$^{1}$, Naveen Kumar Desiraju$^2$, Wolfgang Mack$^{1\dag}$, Emanu\"{e}l A. P. Habets$^{1}$} 
\IEEEauthorblockA{\textit{$^1$ International Audio Laboratories Erlangen\textsuperscript{$\ast$}, Am Wolfsmantel 33, 91058 Erlangen, Germany}
\thanks{\textsuperscript{$\ast$}A joint institution of Fraunhofer IIS and Friedrich-Alexander-Universit{\"a}t Erlangen-N{\"u}rnberg (FAU), Germany.}\\
\IEEEauthorblockA{\textit{$^2$ Fraunhofer IIS, Am Wolfsmantel 33, 91058 Erlangen, Germany}\\}
\IEEEauthorblockA{\small \textit{\{shrishti.saha.shetu, naveen.kumar.desiraju, 
 emanuel.habets\}@iis.fraunhofer.de}, \small \textit{wolfgang.mack@fau.de} } \thanks{$^\dag$ Wolfgang Mack is now with Cisco.}}

}

\maketitle

\begin{abstract}
The successful deployment of deep learning-based acoustic echo and noise reduction (AENR) methods in consumer devices has intensified interest in low-complexity solutions that achieve robust performance in real-world scenarios. In this work, we propose a Kalman filter - deep neural network hybrid method for AENR, which employs a novel channel-wise sampling-based feature reorientation method and time alignment in the latent space for complexity reduction and robust performance. Experimental results show that the proposed method achieves better echo reduction and comparable noise reduction performance to the considered baseline methods with improved generalization capability in many acoustically adverse scenarios, such as non-linear distortions, low-pass filtering, and large acoustical delays.
\end{abstract}

%%%%%%%%%%%%%%%%%%%%%%%%%%%%%%%%%%%%%%%%%%%%%%%%%%%%%%%%%%%%%%%%%%%%%%%%%%%%%
\section{Introduction}
\label{sec:intro}
%%%%%%%%%%%%%%%%%%%%%%%%%%%%%%%%%%%%%%%%%%%%%%%%%%%%%%%%%%%%%%%%%%%%%%%%%%%%%

AENR technology is crucial for modern communication devices, such as smartphones, video conferencing systems, and smart speakers, which require robust AENR solutions that can consistently suppress echo and noise, ensuring clear communication in challenging acoustic scenarios. Although classical adaptive filter-based algorithms \cite{benesty2001advances, gustafsson2002psychoacoustic, valero2015state,valero2018low,kuech2005nonlinear, halimeh2020efficient} are effective in controlled scenarios, their performance degrades in the presence of nonlinear distortions and time-varying acoustic conditions  \cite{franzen2022deep, Halimeh9414868, peng2021acoustic}. Recently, deep neural network (DNN)-based approaches have been widely adopted, offering improved AENR performance, either as part of hybrid systems with adaptive filters \cite{peng2021acoustic, Halimeh9414868,haubner2021synergistic, zhang2023two, mack2023hybrid, franzen2022deep, valin2021low} or in end-to-end systems \cite{braun2022task, indenbom2022deep, indenbom2023deepvqe, zhang2022multi, westhausen2021acoustic}. However, most high-performance solutions are resource-intensive, making them impractical for embedded devices. 

In hybrid systems, adaptive filter-based algorithms are typically employed for linear acoustic echo cancellation, while DNNs suppress residual and nonlinear echo \cite{valin2021low, 9746039, shetu2024aenr, Halimeh9414868}. This division of tasks allows the development of low-complexity DNN post-filters.  In \cite{shetu2024aenr}, 
 we integrated the ULCNet model \cite{shetu2024ultra}, designed initially for noise reduction (NR), into a hybrid AENR system, achieving performance comparable to other evaluated methods. The obtained ULCNet$_{\text{AENR}}$ model demonstrated low computational and memory requirements, making it highly suitable for real-time processing on embedded devices. Despite these advantages, it exhibits sub-optimal performance in certain scenarios due to its high dependence on the convergence state of the Kalman filter (KF). Additionally, the employed channel-wise subband feature reorientation method (C-SubFR) \cite{channelwise} can render the ULCNet model vulnerable in certain adverse conditions, such as lowpass or bandpass-filtered input signals.

To address these limitations, i)~we propose an improved version of the ULCNet$_{\text{AENR}}$ model \cite{shetu2024aenr}, called Align-ULCNet, which integrates two parallel streams for independent encoding of input signals and introduces a cross-attention-based time alignment (TA) block in the latent space. The integration of the TA block improves the system's performance, especially in cases where the preceding adaptive filter has not fully converged. ii)~We introduce a novel channel-wise sampling-based feature reorientation (C-SamFR) method to improve robustness against different input signal characteristics, particularly for band-limited input signals. iii)~We investigate suitable input feature combinations for the proposed model by evaluating their performance in adverse scenarios.

The remainder of this paper is structured as follows. In Section \ref{sec:Proposed}, we present the signal model and the proposed Align-ULCNet model, including the proposed C-SamFR method and the TA block. Section \ref{sec:training} contains details regarding model training. Section \ref{sec:Res} provides performance evaluations for acoustic echo reduction (AER) and NR tasks, comparing our proposed approach with existing methods. Section \ref{sec:ASD} includes ablation studies analyzing the impact of key design choices, followed by a discussion of the results.

%%%%%%%%%%%%%%%%%%%%%%%%%%%%%%%%%%%%%%%%%%%%%%%%%%%%%%%%%%%%%%%%%%%%%%%%%%%%%
\section{Signal Model and Proposed Method}
\label{sec:Proposed}
%%%%%%%%%%%%%%%%%%%%%%%%%%%%%%%%%%%%%%%%%%%%%%%%%%%%%%%%%%%%%%%%%%%%%%%%%%%%%
\begin{figure}[t]
\centering
\includegraphics[width=.47\textwidth]{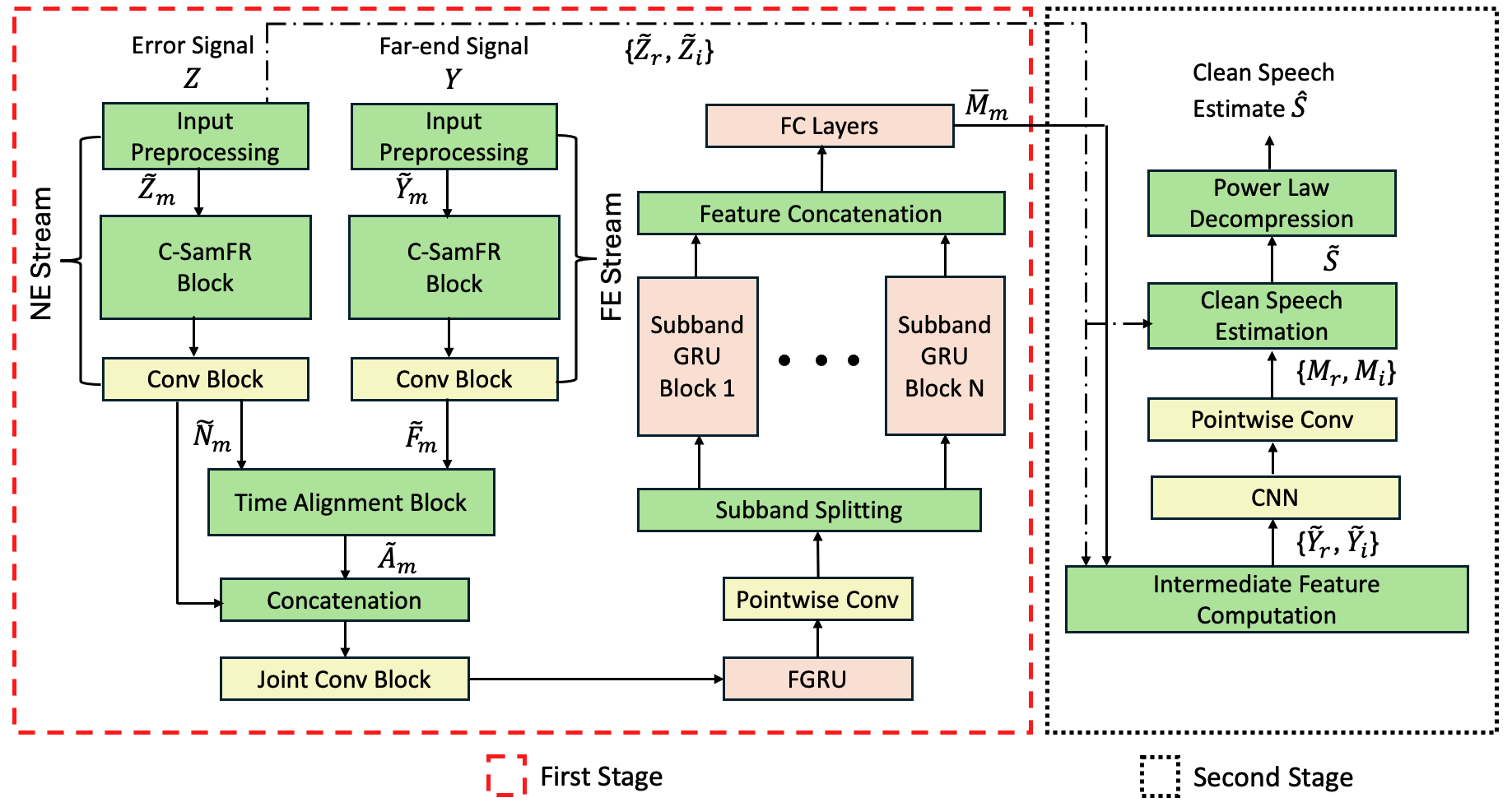}
\vspace{-0.1cm}
\caption{Proposed low-complexity Align-ULCNet model} 
\label{fig:AULC}
\vspace{-0.5cm}
\end{figure}

We consider a duplex-communication scenario, where the far-end signal $y$ is reproduced by a loudspeaker and the microphone signal $x$ is composed of the desired near-end (NE) speech $s$, the acoustic echo $e$ and the background noise $v$:
\vspace{-0.1cm}
\begin{equation}
    x(n) = s(n) + e(n) + v(n),
    \label{eq:micTime}
\vspace{-0.1cm}
\end{equation}
where $n$ denotes the discrete-time sample index. We consider a hybrid system in which the microphone signal is first processed with a KF \cite{kuech2014state}, followed by a DNN-based post-filter. The KF computes an estimate for the linear acoustic echo $\hat{e}(n)=\text{KF}\big( x(n),y(n) \big)$, which is then subtracted from the microphone signal to obtain the error signal $z$, i.e.,

\vspace{-0.1cm}
 \begin{equation}
 z(n)  = x(n) - \hat e(n).
    \label{eq:errorFreq}
\vspace{-0.1cm}
 \end{equation}
The error signal $z$ contains residual echo $r$ and background noise $v$, which are subsequently suppressed using the proposed $\text{Align-ULCNet}$ model. 
%\vspace{0.1cm}

\noindent \textbf{Input Preprocessing:} The input to the proposed Align-ULCNet model is the short-time Fourier transform (STFT) features of the error signal $z$ and far-end signal $y$, as shown in Fig. \ref{fig:AULC}. Subsequently, the power-law compressed magnitude features are obtained in the input preprocessing block, as described in \cite{shetu2024ultra}.

\noindent \textbf{Align-ULCNet:}  The proposed model has two key modifications to the two-stage ULCNet$_{\text{AENR}}$ model \cite{shetu2024aenr}, i)~the convolutional encoder in the first stage has been redesigned by introducing two parallel convolutional streams — NE and FE — to process the two input signals separately, obtaining the encoded NE and FE features $\widetilde{N}_\textrm{m}$ and $\widetilde{F}_\textrm{m}$, respectively. This parallel structure ensures independent extraction of near-end and far-end features.  ii)~A cross-attentional TA block is then integrated between the NE and FE  streams to align their features in the latent space. The TA block jointly processes $\widetilde{N}_\textrm{m}$ and $\widetilde{F}_\textrm{m}$  to generate time-aligned FE features,  represented as $\widetilde{A}_\textrm{m}$. These aligned features are concatenated with NE features $\widetilde{N}_\textrm{m}$ along the channel dimension 
and further processed in the joint Conv block. The final complex mask $M$ is obtained by processing the convolutional features through FGRU, subband GRUs, FC layers, and CNN blocks. This complex mask $M$ is then used to reconstruct the power-law compressed \cite{shetu2024ultra} near-end speech estimate $\tilde{S}$ as follows:
\vspace{-0.1cm}
   \begin{equation}
        \widetilde{S} = \widetilde{Z}_{\textrm{m}} \cdot M_{\textrm{m}} \cdot e^{j(\widetilde{Z}_{\textrm{p}} + M_{\textrm{p}})},
        \label{eq:CleanSpeechEst}
\vspace{-0.1cm}
    \end{equation}
where $j=\sqrt{-1}$, $M_{\textrm{m}}$ and $M_{\textrm{p}}$ represent the magnitude and phase components of the complex-valued mask $M$, and $\widetilde{Z}_{\textrm{m}}$ and $\widetilde{Z}_{\textrm{p}}$ represent the power-law compressed magnitude and phase components of the error signal $Z$, respectively.
%\vspace{0.1cm}

\begin{figure}[t]
\centering
\includegraphics[width=.45\textwidth]{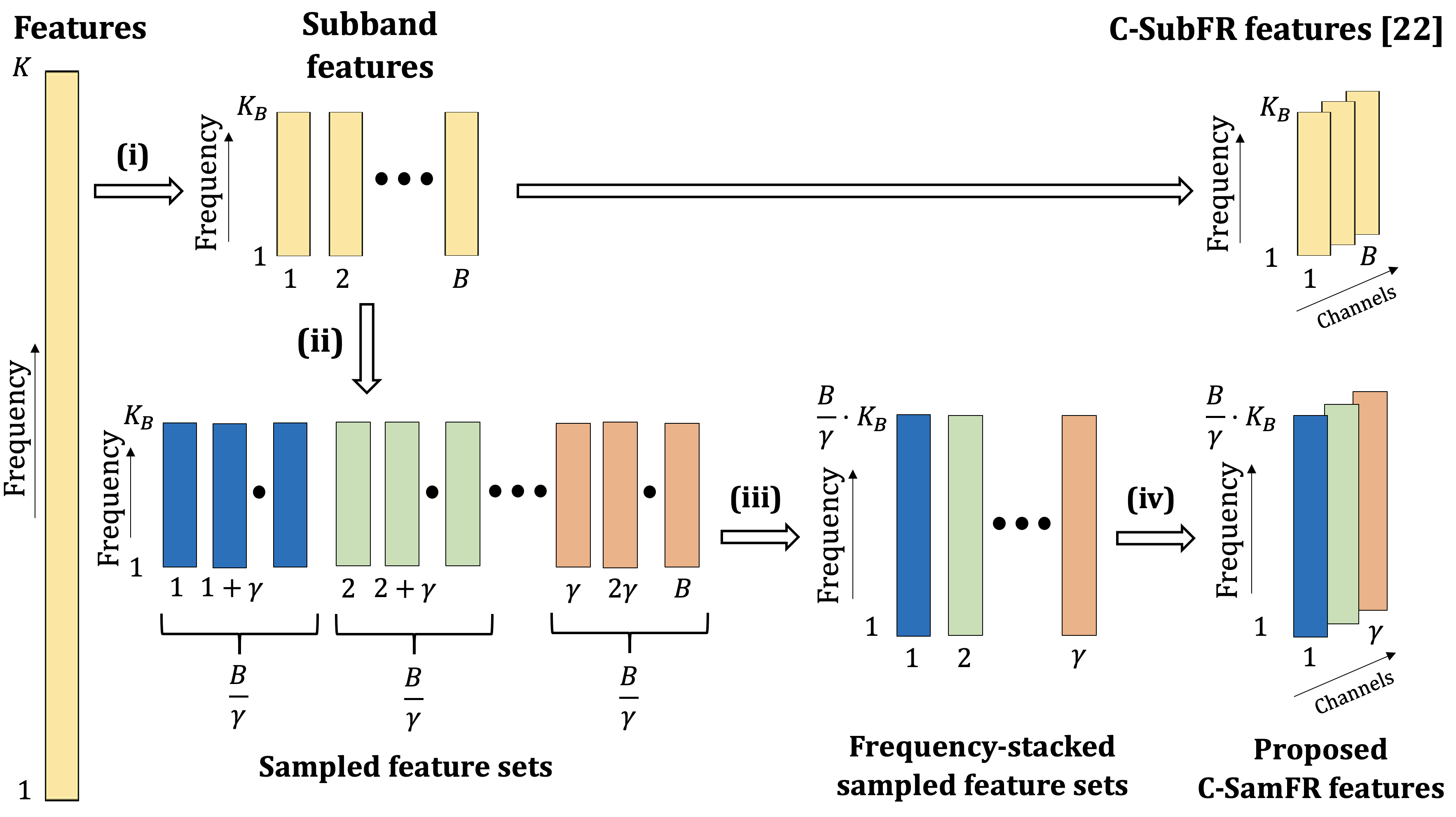}
\vspace{-0.1cm}
\caption{Proposed C-SamFR method} 
\label{fig:TSD}
\vspace{-0.5cm}
\end{figure}

\noindent \textbf{C-SamFR method:} The C-SamFR method rearranges input features in a non-sequential manner along the frequency dimension as shown in Fig. \!\ref{fig:TSD}. The C-SamFR method is applied separately to the $K$-element power-law compressed magnitude features $\widetilde{Z}_\text{m}$ and $\widetilde{Y}_\text{m}$, involving four different steps, (i) for each input signal in each frame, $K$ features are divided into $B$ subbands, each containing $K_B$ contiguous frequency bins, with an overlap factor of $0 \leq \beta < 1$ \cite{channelwise}. (ii) These $B$ subbands are then rearranged using a sampling factor $\gamma$, forming sets containing $\frac{B}{\gamma}$ subbands each. (iii) The subbands in each set are then stacked along the frequency dimension to generate a total of $\gamma$ sampled feature sets. (iv) Finally, these feature sets are stacked along the channel dimension to obtain the final C-SamFR features. The resulting C-SamFR features enhance feature representation compared to the C-SubFR features (upper part of Fig. \ref{fig:TSD}) by reducing the likelihood of subbands containing only zeros in the channel dimension.
%\vspace{0.1cm}

\noindent \textbf{Time Alignment:} To reduce computational complexity and leverage convolutional features for soft alignment, we implement a convolutional cross-attentional TA block in the latent space. This processing was inspired by the approach in \cite{indenbom2023deepvqe}.  The NE features $\widetilde{N}_\textrm{m} \in \mathbb{R}^{L \times T \times P}$ and FE features $\widetilde{F}_\textrm{m} \in \mathbb{R}^{L \times T \times P}$ (where $L$ denotes the number of channels, $T$ denotes the number of frames in the time dimension, and $P$ denotes the number of features in the feature dimension) are first transformed by point-wise convolution layers into $N \in \mathbb{R}^{H \times T \times P}$ and $F \in \mathbb{R}^{H \times T \times P}$, respectively, where $H$ is the number of similarity channels. Then, $F$ is unfolded along the time axis to produce delayed features $F_u \in \mathbb{R}^{H \times T \times D_{\text{max}} \times P}$, where $D_{\text{max}}$ is the maximum echo delay. A dot product between $N$ and $F_u$ along the feature axis yields $C \in \mathbb{R}^{H \times T \times D_{\text{max}}}$:
%\vspace{-0.3cm}
\vspace{-0.5em}
\begin{equation}
C(h, t, d) = \sum_{p=1}^{P} N(h, t, p) \cdot F_u(h, t, d, p),
\vspace{-0.2cm}
\end{equation}
which is further processed by a convolutional layer to obtain the delay probability distribution $D \in \mathbb{R}^{T \times D_{\text{max}}}$ using a softmax along the delay dimension $d$. Finally, the aligned FE features $\widetilde{A}_\textrm{m} \in \mathbb{R}^{H \times T \times P}$ are computed as a weighted sum of $F_u$ with the delay probabilities from $D$ along the time axis:
%\vspace{-0.2cm}
\vspace{-0.5em}
\begin{equation}
\widetilde{A}_\textrm{m}(h,t, p) = \sum_{d=1}^{D_{\text{max}}} D(t, d) \cdot F_u(h, t, d, p).
\vspace{-0.1cm}
\end{equation}

%%%%%%%%%%%%%%%%%%%%%%%%%%%%%%%%%%%%%%%%%%%%%%%%%%%%%%%%%%%%%%%%%%%%%%%%%%%%%
%\vspace{-0.2cm}
%\section{Experiments and Results}
%\label{sec:Exp}
%%%%%%%%%%%%%%%%%%%%%%%%%%%%%%%%%%%%%%%%%%%%%%%%%%%%%%%%%%%%%%%%%%%%%%%%%%%%%
\textbf{\begin{table*}
\centering
\small
\caption{AECMOS \cite{purin2022aecmos} results on AEC Challenge blind test sets.}
\begin{tabular} {l c c c c c c c c c c }
    \toprule
    \multicolumn{1}{l}{}                                & \multicolumn{2}{c}{Computational}     & \multicolumn{4}{c}{Interspeech 2021 \cite{cutler2021interspeech}} & \multicolumn{4}{c}{ICASSP 2023 \cite{cutler2024icassp}} \\
    \cmidrule(lr){4-7} \cmidrule(lr){8-11}
                                                        & \multicolumn{2}{c}{Complexity}        & \multicolumn{2}{c}{DT}        & FST           & NST           & \multicolumn{2}{c}{DT}        & FST           & NST           \\
    \cmidrule(lr){2-3} \cmidrule(lr){4-5} \cmidrule(lr){6-6} \cmidrule(lr){7-7} \cmidrule(lr){8-9} \cmidrule(lr){10-10} \cmidrule(lr){11-11}
    \textbf{Processing$^1$}                                 & Params [M]        & GMACS             & EMOS          & DMOS          & EMOS          & DMOS          & EMOS          & DMOS          & EMOS          & DMOS          \\
    \midrule
    Peng et al.* \cite{peng2021acoustic}                 & 10.20             & 2.52              & 4.36          & 4.23          & 4.34          & 4.26          & -             & -	            & -             & -             \\
    Braun et al. \cite{braun2022task}                   & -                 & -                 & 4.55          & \textbf{4.25} & 4.35          & 4.18          & -             & -	            & -             & -             \\    
    Zhang et al.* \cite{zhang2023two}                    & 9.56              & -                 & -             & -             & -             & -             & \textbf{4.72} & 4.16          & 4.70          & -          \\
    Deep-VQE* \cite{indenbom2023deepvqe}                 & 7.50              & 4.02              & -             & -             & -             & -             & 4.70          & \textbf{4.29} & 4.69          & - \\
     Align-CRUSE \cite{indenbom2022deep}                 & 0.74              & -                 & 4.45          & 4.07          & 4.67          & -             & 4.60          & 3.95          & 4.56          & -             \\
    %\hline
    ULCNet$_{\text{AENR}}$(C-SubFR) \cite{shetu2024aenr}         & \textbf{0.69}     & \textbf{0.10}     & 4.61          & 3.79          & 4.64          & 4.28          & 4.54          & 3.58          & 4.73          & 4.15          \\
    ULCNet$_{\text{AENR}}$(C-SamFR)                        & \textbf{0.69}     & \textbf{0.10}     & 4.58          & 3.93          & 4.53          & \textbf{4.32} & 4.58          & 3.74          & 4.71          & 4.15          \\
    \textbf{Proposed Align-ULCNet}                      & \textbf{0.69}     & \textbf{0.10}     & \textbf{4.66} & 3.95          & \textbf{4.75} & 4.29          & 4.60          & 3.80          & \textbf{4.77} & \textbf{4.28}          \\
    \bottomrule
\end{tabular}
\label{tab:AER}
\vspace{-1em}
\end{table*}}

\vspace{-1em}
\section{Experiments and Results}
\label{sec:Exp}
%\vspace{-0.6cm}
%%%%%%%%%%%%%%%%%%%%%%%%%%%%%%%%%%%%%%%%%%
\subsection{Experimental Design}
\label{sec:training}
%%%%%%%%%%%%%%%%%%%%%%%%%%%%%%%%%%%%%%%%%%
\noindent \textbf{Model Parameters}: The Conv blocks in the NE and FE streams comprised two separable convolution layers, each with $L = 32$ filters and stride of $(1 \times 1)$, and with kernel sizes of $(1 \times 5)$ and $(1 \times 3)$, respectively. Downsampling by a factor of 2 was performed through max-pooling along the frequency dimension. In the TA block, we employed point-wise convolutions with $H = 32$ filters and used a kernel size of $(5 \times 3)$ for the convolutional operation on the dot product features $C$. The Joint Conv block included two convolution layers with 64 and 96 filters, a kernel size of $(1 \times 3)$, and a stride of $(1 \times 2)$. The remaining blocks in the Align-ULCNet model follow the same parameterization as described in \cite{shetu2024ultra}.

%\vspace{0.1cm}

\noindent \textbf{Experimental Parameters}: The KF used a recursive Kalman gain of 0.8 and 10 partition blocks, as mentioned in \cite{kuech2014state, haubner2021synergistic}. We chose $N_{\text{FFT}}=512$, a window length of 32 ms, and a hop size of 16 ms such that $K = 257$. We used a power-law compression factor $\alpha=0.3$ for the input preprocessing step. For the C-SamFR method, we used $K_B=2$ with an overlap factor of $\beta=0$, such that $B = 130$, and a sampling factor of $\gamma=5$, while for the TA block, we used $D_{\text{max}} = 64$, which roughly corresponded to a maximum echo delay of 1~s. For training, we used the Adam optimizer with an initial learning rate of 0.004, which decayed by a factor of 10 when the validation loss did not improve with a patience of one epoch. Each training sample was of 3~s duration, a batch size of 64 was chosen, and the model was trained for 20k steps per epoch.
%\vspace{0.1cm}

\noindent \textbf{Training Dataset}: We used the measured echo and far-end signals provided in \cite{cutler2024icassp}, as well as clean and noisy speech signals from \cite{ reddy2020interspeech} to create a training dataset of 1100 hours at 16 kHz sampling rate using the methodology mentioned in \cite{shetu2024aenr}. Different scenarios for the microphone signal, e.g., near-end single-talk (NST), far-end single-talk (FST), and double-talk (DT), were simulated following the methods proposed in \cite{mack2023hybrid}, with SNR $\in [-5,30]$ and SER $\in [-20,20]$ dB.
%\vspace{0.1cm}

\noindent \textbf{Evaluation Dataset and Metrics}: The Interspeech 2021 \cite{cutler2021interspeech} and ICASSP 2023 \cite{cutler2024icassp} AEC Challenge datasets are used for evaluating the AER performance, and the DNS challenge 2020 dataset \cite{reddy2020interspeech} is used for evaluating the NR performance. We use AECMOS \cite{purin2022aecmos}, SI-SDR \cite{le2019sdr}, BAKMOS, and SIGMOS \cite{reddy2022dnsmos} as the evaluation metrics. 

\blfootnote{$^1$Metrics reported for SOTA methods are taken directly from their respective publications, with an asterisk (*) indicating results reported as part of the AEC challenge.} 

\begin{table}
\small
\centering
\caption{Objective results for NR on DNS Challenge \cite{reddy2020interspeech} non-reverb test set}
\begin{tabular} {l c c c }
    \toprule    
    \textbf{Processing}                                 & SI-SDR            & SIGMOS        & BAKMOS        \\
    \midrule
    Noisy                                               & 9.06              & 3.39          & 2.62          \\
    DeepFilterNet \cite{schroter2022deepfilternet}      & 16.17             & 3.49          & 4.03          \\
    DeepFilterNet2 \cite{schroter2022deepfilternet2}    & 16.60             & \textbf{3.51} & \textbf{4.12} \\
    ULCNet$_{\text{MS}}$\cite{shetu2024ultra}           & 16.34             & 3.46          & 4.06          \\
    ULCNet$_{\text{Freq}}$\cite{shetu2024ultra}         & \textbf{16.67}    & 3.38          & 4.09          \\
    ULCNet$_{\text{AENR}}$ \cite{shetu2024aenr}         & 15.58             & 3.30          & 4.05          \\
    \textbf{Proposed  Align-ULCNet}                     & 16.12             & 3.33          & 4.08          \\
    \bottomrule
\end{tabular}
\label{tab:ObjectiveresultNR}
\end{table}

\begin{table}
\small
\centering
\caption{EMOS results for the proposed Align-ULCNet model for different input combinations}
\begin{tabular}{c c c c c}
    \toprule
                                            & \multicolumn{2}{c}{Input Signals}             & \multicolumn{2}{c}{EMOS} \\
    \cmidrule(lr){2-3} \cmidrule(lr){4-5}
    TA block                                & NE stream             & FE stream             & DT            & FST \\
    \midrule
    KF                                      & -                     & -                     & 1.73          & 2.19 \\
    No                                      & $Z$                   & $Y$                   & 3.73          & 3.91 \\
    \textbf{Yes}                            & \boldmath{$Z$}        & \boldmath{$Y$}        & \textbf{4.20} & \textbf{4.69} \\
    Yes                                     & $Z$ and $\widehat{E}$ & $Y$                   & 4.04          & 4.18 \\
    Yes                                     & $Z$                   & $Y$ and $\widehat{E}$ & 3.99          & 4.39 \\
    \bottomrule
\end{tabular}
\label{tab:InputSignals}
\vspace{-0.3cm}
\end{table}
% TO-DO: The inputs for the non-TA model are incorrect. Currently they have been reported as Z and Y. In reality they are Z+E and Y. We need to train a new model with Z and Y as inputs without the TA module after review. [No time to retrain. Wouldn't make much difference anyway]
\vspace{-0.1cm}
%%%%%%%%%%%%%%%%%%%%%%%%%%%%%%%%%%%%%%%%%%

\subsection{Results}
\label{sec:Res}
%%%%%%%%%%%%%%%%%%%%%%%%%%%%%%%%%%%%%%%%%%
\noindent \textbf{AER Performance}: Table \ref{tab:AER} presents a comparison of the echo reduction performance of the proposed Align-ULCNet model against six recent baseline methods from literature, as well as the ULCNet$_{\text{AENR}}$ model combined with the C-SamFR methods. The results demonstrate that the proposed model achieves substantial improvements over both versions of ULCNet$_{\text{AENR}}$ across all objective metrics. The enhancement in EMOS metrics can be attributed to the integration of the TA block, while the improvement in DMOS metrics primarily results from using the C-SamFR method. Additionally, our approach surpasses the baseline methods in EMOS metrics while delivering slightly lower DT DMOS scores, but at a significantly reduced computational cost and memory footprint.

\noindent \textbf{NR Performance}: In Table \!\ref{tab:ObjectiveresultNR}, we compare the NR performance of the proposed Align-ULCNet model against five different low-complexity dedicated NR methods. While DeepFilterNet2 \cite{schroter2022deepfilternet2} achieves slightly higher SIGMOS, our method maintains comparable NR performance (BAKMOS) with significantly lower computational complexity \cite{shetu2024ultra}. This suggests that Align-ULCNet effectively balances echo suppression and noise reduction without excessive parameter overhead. In informal listening tests, we observed that the perceptual quality of the Align-ULCNet is comparable to the other methods under test. The processed samples can be found here: \url{https://fhgainr.github.io/alignulcnet-aenr/}.

\begin{figure}[t]
    \centering
    \begin{subfigure}{0.46\linewidth}
        \includegraphics[width=\linewidth]{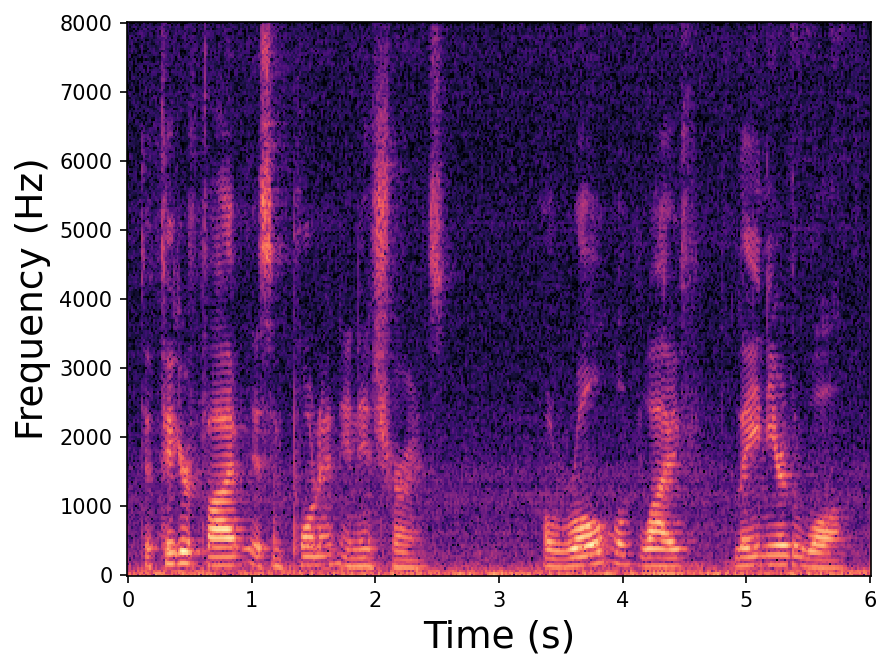}
        \caption{\small Wideband signal}
        \label{fig:sub1}
    \end{subfigure}
    \begin{subfigure}{0.46\linewidth}
        \includegraphics[width=\linewidth]{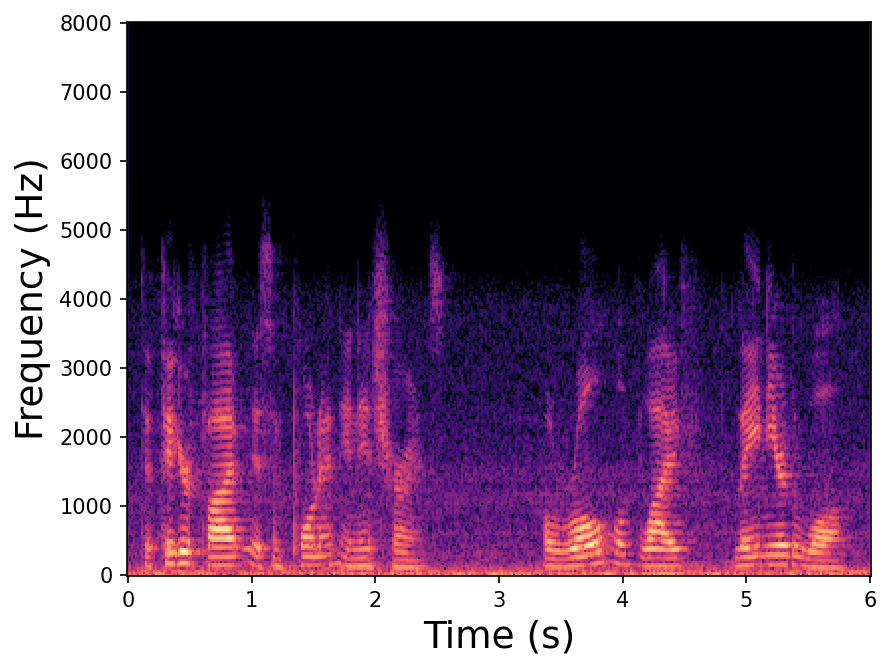}
        \caption{\small Band-limited at 4kHz}
        \label{fig:sub2}
    \end{subfigure}
    \\
    \begin{subfigure}{0.46\linewidth}
        \includegraphics[width=\linewidth]{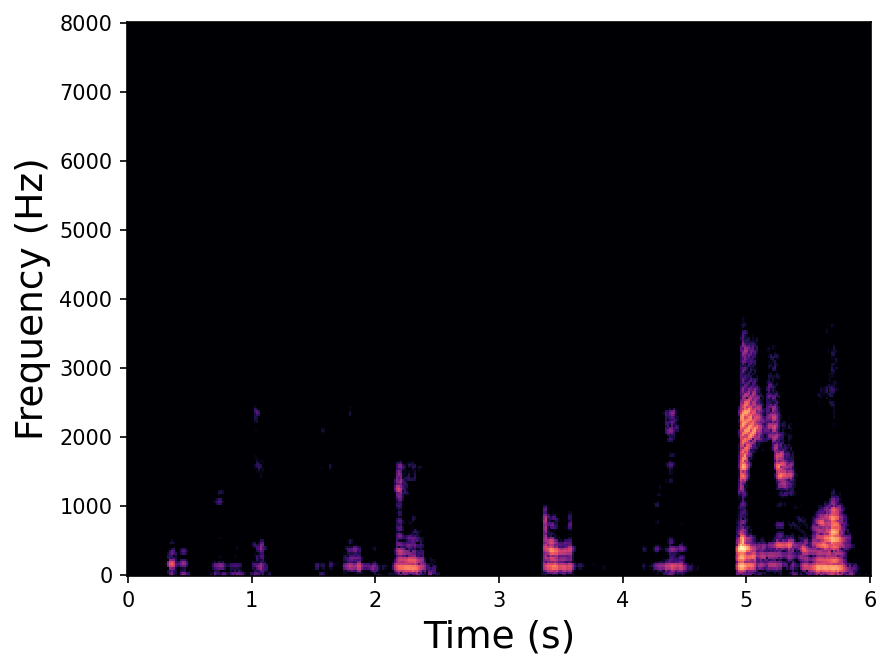}
        \caption{\small ULCNet$_{\text{AENR}}$(C-SubFR)}
        \label{fig:sub3}
    \end{subfigure}
    \begin{subfigure}{0.46\linewidth}
        \includegraphics[width=\linewidth]{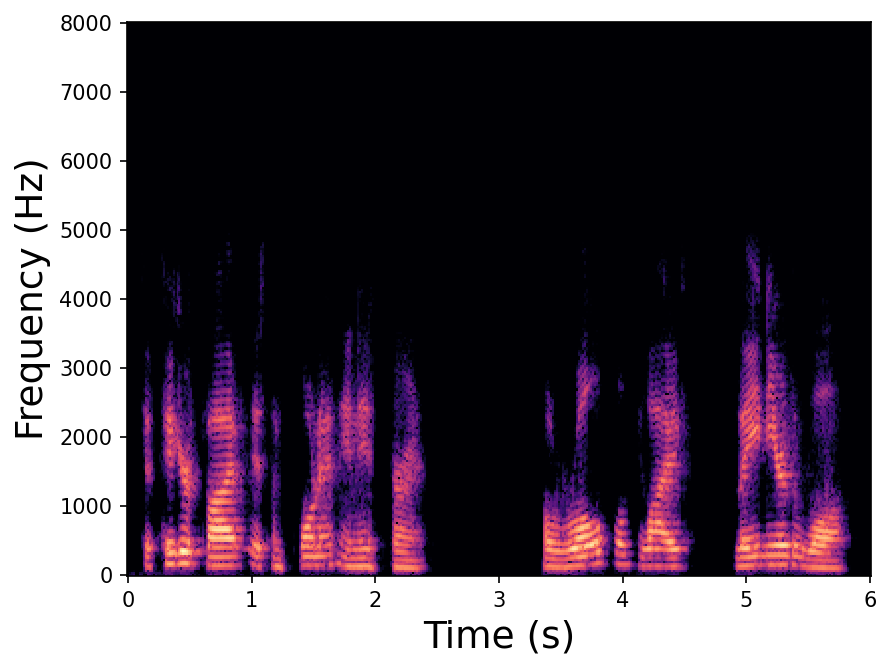}
        \caption{\small   ULCNet$_{\text{AENR}}$(C-SamFR)}
        \label{fig:sub4}
    \end{subfigure}
    \caption{Example of (a) a noisy NST wideband signal, (b) band-limited at 4 kHz frequency, output of  ULCNet$_{\text{AENR}}$ model \cite{shetu2024aenr} using (c)  C-SubFR \cite{channelwise} and (d) proposed C-SamFR method.}
    \label{fig:CSubFeature}
    \vspace{-2em}
\end{figure} 
%%%%%%%%%%%%%%%%%%%%%%%%%%%%%%%%%%%%%%%%%%
\subsection{Ablation Study and Discussion}
\label{sec:ASD}

%%%%%%%%%%%%%%%%%%%%%%%%%%%%%%%%%%%%%%%%%%
\noindent \textbf{Effect of C-SamFR Method:} The C-SubFR method \cite{channelwise}, as integrated in the ULCNet$_{\text{AENR}}$ \cite{shetu2024aenr} model, exhibits performance issues when the input signal is band-limited. This is due to the sequential subband splitting process, which can result in one or more subbands being completely filled with zeros in the channel dimension, thereby negatively impacting the model's overall performance. Our proposed C-SamFR method addresses this issue by ensuring that no subbands contain only zeros in the channel dimension. As shown in Fig. ~\!\!\ref{fig:CSubFeature}, in the NST scenario with the microphone signal band-limited at 4kHz, the (d) ULCNet$_{\text{AENR}}$(C-SamFR) method maintains satisfactory performance, whereas the (c) ULCNet$_{\text{AENR}}$(C-SubFR) model tends to suppress the NE signal excessively. The proposed C-SamFR method also significantly enhances speech quality compared to the C-SubFR method, as reflected in the DMOS metrics in Table \ref{tab:AER}, while maintaining similar echo reduction performance, as indicated by the EMOS metrics.

\noindent \textbf{Different Input Combinations:} 
In this work, we also examine the impact of various input signal combinations (the STFT-domain error signal $Z$, echo estimate $\widehat{E}$ and far-end signal $Y$) on the Align-ULCNet model performance for the scenarios where the KF fails to converge. For this experiment, we used the same AEC challenge test datasets as mentioned in Table \ref{tab:AER}, and selected the samples where the KF failed to perform satisfactorily ($\approx 60$ samples). We can observe the results in Table \ref{tab:InputSignals}, where the first row contains the scores obtained after KF processing. The results clearly demonstrate that our proposed Align-ULCNet model, which takes the error signal $Z$ in the NE stream and the far-end signal $Y$ in the FE stream, outperforms all other input signal combinations.
%\vspace{0.1cm}

\begin{figure}[t]
\centering
\includegraphics[width=.5\textwidth]{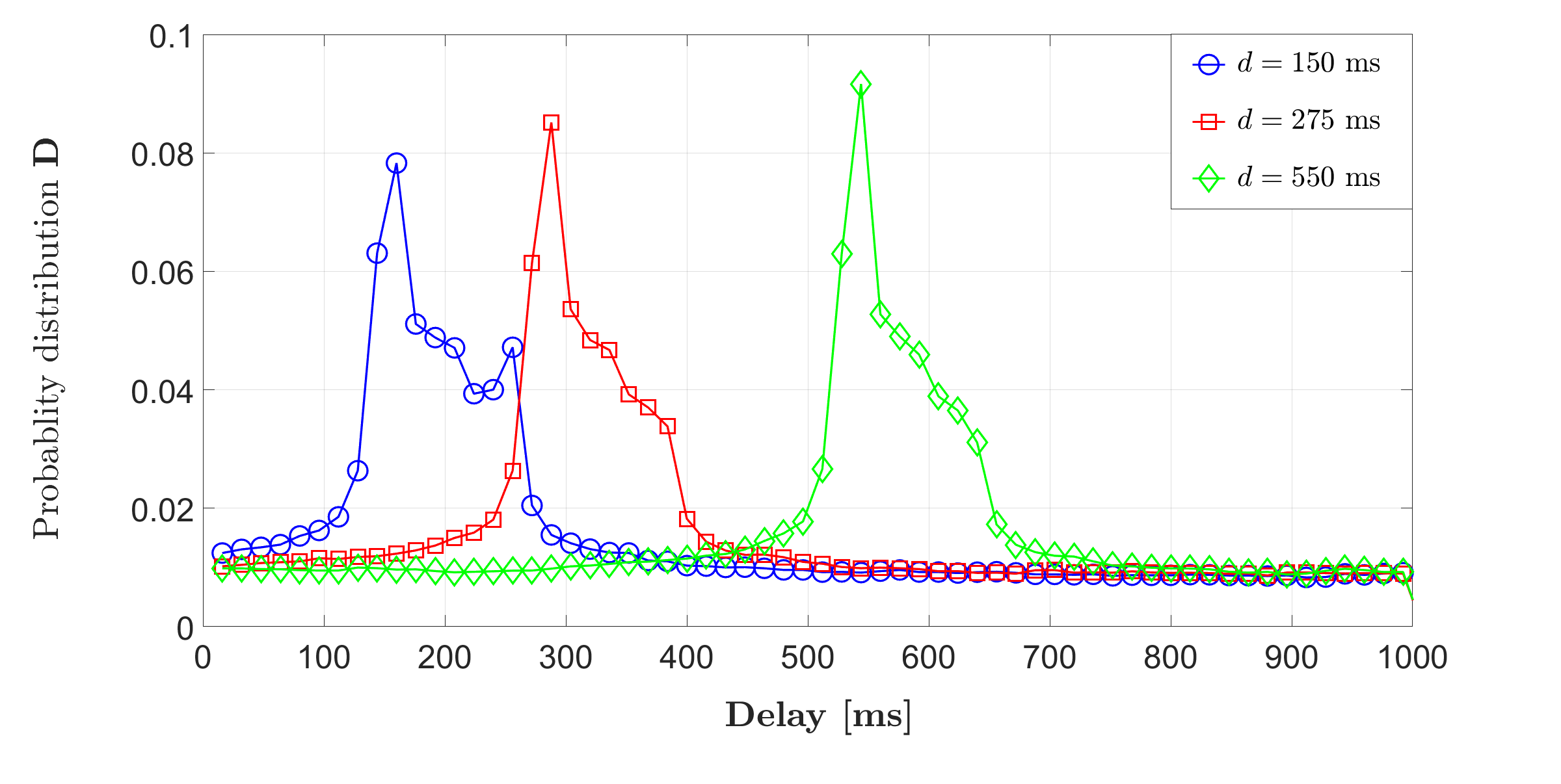}
\vspace{-2em}
\caption{Estimated delay probability distribution $\mathbf{D}$ for different amounts of acoustic delay $d$}
\label{fig:TA}
\vspace{-2em}
\end{figure}

\noindent \textbf{Analysis of Time Alignment:} The experiment with different input combinations further highlights the effectiveness of the TA block. As shown in Table~\ref{tab:InputSignals}, omitting the TA block with the same input combination ($Z$ and $Y$) significantly reduces performance, with EMOS dropping by $-0.47$ and $-0.78$ for DT and FST scenarios, respectively. This supports the hypothesis that the TA block reduces the system's reliance on the KF. Figure \ref{fig:TA} shows the accurate delay probability distribution $\mathbf{D}$ computed by the TA block for varying acoustic delays. These probabilities are used in the weighted sum computation of time-aligned FE features $\widetilde{A}_\textrm{m}$, highlighting the need for a sufficiently long delay buffer in real-time streaming applications. 

%\vspace{0.1cm}
\noindent \textbf{Discussion:} Our proposed Align-ULCNet model, while being computationally lightweight, delivers competitive AER and on-par NR results as compared to the baseline AER and dedicated low-complexity NR methods. We demonstrate that the proposed C-SamFR method ensures robust performance, particularly when the input signal is band-limited - a scenario likely in mass consumer device deployments. Including the TA block in the Align-ULCNet model enhances its independence from the KF, further improving overall performance under adverse conditions. Although objective metrics such as DMOS and SIGMOS indicate slightly inferior results with respect to speech quality, previous studies \cite{shetu2024ultra,shetu2024aenr} have shown that the use of power-law compression can negatively impact these metrics. However, in informal listening tests, the results of the proposed model remain preferable. When implemented efficiently, our model requires only a small buffer for storing the far-end features from previous frames, achieving a real-time factor of $\approx$16$\%$ on a Cortex-A53 1.43 GHz processor.

%%%%%%%%%%%%%%%%%%%%%%%%%%%%%%%%%%%%%%%%%%%%%%%%%%%%%%%%%%%%%%%%%%%%%%%%%%%%%
\section{Conclusion}
\label{sec:Con}
%%%%%%%%%%%%%%%%%%%%%%%%%%%%%%%%%%%%%%%%%%%%%%%%%%%%%%%%%%%%%%%%%%%%%%%%%%%%%
We proposed a low-complexity hybrid approach for robust AENR. Our proposed Align-ULCNet model achieves objective results that are superior or comparable to other evaluated methods for both echo and noise reduction tasks. With a small model size of only 0.69M parameters and a computational complexity of 0.10 GMACS, the proposed method is well-suited for deployment in resource-constrained consumer devices. Additionally, we provide insights into the various processing blocks within the model and
their contributions to the overall performance of the proposed
model. 

\pagebreak
\section{Acknowledgment}
This work has been supported by the Free State of Bavaria in the DSAI project. 

% -------------------------------------------------------------------------
%\footnotesize
\let\oldbibliography\thebibliography
\renewcommand{\thebibliography}[1]{%
  \oldbibliography{#1}%
  %\footnotesize
  \setlength{\itemsep}{1pt}%
}
\bibliographystyle{IEEEbib}
\bibliography{strings}

\end{document}